\documentclass[12pt,preprint]{aastex}
\usepackage{mkfig}

\begin{document}

\title{\bf New Mass Loss Measurements from Astrospheric Ly$\alpha$
  Absorption}

\author{B. E. Wood\altaffilmark{1},
  H. -R. M\"{u}ller\altaffilmark{2,3}, G. P. Zank\altaffilmark{3},
  J. L. Linsky\altaffilmark{1}, S. Redfield\altaffilmark{4}}

\altaffiltext{1}{JILA, University of Colorado and NIST, Boulder, CO
  80309-0440; woodb@origins.colorado.edu, jlinsky@jila.colorado.edu.}
\altaffiltext{2}{Department of Physics and Astronomy, Dartmouth College,
  6127 Wilder Lab, Hanover, NH 03755-3528; Hans.Mueller@Dartmouth.edu.}
\altaffiltext{3}{Institute of Geophysics and Planetary Physics,
   University of California at Riverside, 1432 Geology, Riverside, CA 92521;
   zank@ucrac1.ucr.edu.}
\altaffiltext{4}{Harlan J. Smith Postdoctoral Fellow, McDonald Observatory,
  University of Texas, Austin, TX 78712-0259; sredfield@astro.as.utexas.edu.}

\begin{abstract}

     Measurements of stellar mass loss rates are used to assess
how wind strength varies with coronal activity and age for solar-like
stars.  Mass loss generally increases with activity, but we find
evidence that winds suddenly weaken at a certain activity threshold.
Very active stars are often observed to have polar starspots, and we
speculate that the magnetic field geometry associated with these
spots may be inhibiting the winds.  Our inferred mass-loss/age relation
represents an empirical estimate of the history of the solar wind.
This result is important for planetary studies as well as
solar/stellar astronomy, since solar wind erosion
may have played an important role in the evolution of planetary
atmospheres.

\end{abstract}


\section{INTRODUCTION}

     The weak winds generated by solar-like stars are normally undetectable
to remote sensing.  However, these winds ultimately collide with the
interstellar medium (ISM) surrounding the star, and if the surrounding
ISM is at least partially neutral this collision yields a population
of hot hydrogen atoms that produces a detectable absorption signature
in spectra of the stellar Ly$\alpha$ line.  Spectra obtained by the
{\em Hubble Space Telescope} (HST) have provided detections of this
absorption from hydrogen in the outer regions of our own heliosphere, as
well as many detections of absorption from the ``astrospheres'' surrounding
the observed stars \citep[e.g.,][]{jll96,bew96,bew00}.

     The detection of astrospheres not only represents the
first clear detection of solar-like winds from other stars, it also
allows the first estimates of mass loss rates from these stars, since the
amount of absorption is correlated with the strength of the wind.
These measurements have been used to investigate how mass loss varies
with age and coronal activity for solar-like stars.  Initial results suggest
that younger stars with more active coronae have stronger winds
\citep[][hereafter Paper~1]{bew02}, implying that the solar wind was
stronger in the past.
We have recently analyzed all appropriate Ly$\alpha$ spectra
in the HST archive to search for new astrospheric detections
\citep[][hereafter Paper~2]{bew05}.  In this paper, we estimate mass
loss rates for the seven new astrospheric absorption detections resulting
from this archival Ly$\alpha$ work, and we reassess what the astrospheric
absorption is telling us about how winds
correlate with stellar age and activity.

\section{NEW MASS LOSS MEASUREMENTS}

     Table~1 lists both the new and old mass loss measurements from
astrospheric absorption detections.
In cases where members of a binary system are close enough to be within the
same astrosphere (e.g., $\alpha$~Cen), the spectral types of both stars are
given, since the measured mass loss rate will be the combined mass loss
from both stars.  Likewise, the stellar surface area listed in the table
will in those cases be the binary's combined surface area.  The surface
areas are computed from stellar radii listed in Paper~2.  The coronal X-ray
luminosities listed in the second-to-last column of Table~1
(in ergs~s$^{-1}$) are {\em ROSAT} All-Sky Survey measurements (see
Paper~2).

     In order to measure a mass loss rate from the observed astrospheric
absorption, it is necessary to know the ISM wind velocity seen by the
star ($V_{ISM}$) and the orientation of the astrosphere relative to our
line of sight ($\theta$), which are both listed in Table~1.  The orientation
angle, $\theta$, is the angle between the upwind direction of the ISM flow
seen by the star and our line of sight to the star.  The $V_{ISM}$ and
$\theta$ values in Table~1 are computed from the known proper motions and
radial velocities of the observed stars, and by assuming that the Local
Interstellar Cloud (LIC) flow vector of \citet{rl92}
applies for the ISM surrounding all observed stars.  For a few stars known
to be in directions where the slightly different G cloud vector
of \citet{rl92} is more applicable ($\alpha$~Cen, 36~Oph, 70~Oph), we
use this vector instead.

     The mass loss measurement process is described in detail
in Paper~1 and in \citet{bew04}.  Briefly, hydrodynamic models of the
astrospheres, constrained by the $V_{ISM}$ values in Table~1, are computed
using a four-fluid code developed to model the heliosphere and reproduce
the observed heliospheric Ly$\alpha$ absorption \citep{gpz96,bew00}.
Models with different mass loss rates are computed by varying the stellar
wind density.  Predicted astrospheric Ly$\alpha$ absorption can be computed
from these models for the observed line of sight defined by the $\theta$
value in Table~1.  Paper~1 and \citet{bew04} discuss systematic errors
in detail, such as uncertainties in ISM properties, wind velocities, and
wind variability; concluding that the derived mass loss rates have
uncertainties of order a factor of
two due to these systematics.  Uncertainties of this size are consistent
with the results of \citet{vvi02} and \citet{vf04}, who have studied how
heliospheric Ly$\alpha$ absorption should change with variations in ISM
parameters such as ionization fraction and magnetic field strength.
The assumption of the same 400 km~s$^{-1}$ wind speed in all astrospheric
models (akin to solar low speed streams) is another major source of
uncertainty, its justification being that one might expect similar wind
speeds from stars with similar spectral types and surface escape
speeds \citep{bew04}.

     Figure~1 compares the observed astrospheric absorption with model
predictions.  Mass loss rates are quoted relative to the solar value of
$\dot{M}_{\odot}\approx 2\times 10^{-14}$ $M_{\odot}$~yr$^{-1}$.
Table~1 lists the mass loss rates that yield the best fits to
the data.  When evaluating how well a model fits the data, it is more
important that the model fits well near the base of the H~I
absorption than elsewhere, since discrepancies farther from the core of the
absorption can often be resolved simply by altering the assumed stellar
emission profile.  Thus, the $\dot{M}=1\dot{M}_{\odot}$ model for EV~Lac
is deemed a better fit than the $\dot{M}=2\dot{M}_{\odot}$ model,
for example.

     The astrospheric models that yield the best fits to the data
are illustrated in Figure~2.  Most of the absorption we observe comes
from the ``hydrogen wall'' region in between the astropause and the
stellar bow shock, which is colored a purplish-red in Figure~2.
Because it will generally take more than a decade for wind material to
reach these distances \citep[e.g.,][]{gpz96}, mass loss rate measurements
from astrospheric absorption will typically be indicative of the average
mass loss over decadal timescales, except for the most compact astrospheres
($\epsilon$~Ind, 61~Cyg~A, and DK~UMa), and the astrospheric absorption
will not vary on shorter timescales such as those associated with activity
cycles.  We also note that in cases where the size of the
modeled astrosphere indicates that both stars of a binary system are
within the same astrosphere ($\alpha$~Cen, 36~Oph, $\lambda$~And, 70~Oph,
and $\xi$~Boo), there is no way for us to tell how much each star is
contributing to the combined binary wind or if the combined wind's ram
pressure has been reduced by wind interaction effects.

     For one of the new astrospheric detections,
HD~128987, we find that the astrospheric models are unable to adequately
fit the data regardless of the assumed mass loss rate, so no mass loss
measurement is listed for this star in Table~1.  The primary cause of our
difficulties with HD~128987 is its very low ISM speed of
$V_{ISM}=8$ km~s$^{-1}$.  High ISM velocities yield more heating and
deceleration of H~I at the stellar bow shock.  This results in astrospheric
H~I that is hot and highly decelerated, meaning that Ly$\alpha$ absorption
from this material is broad and shifted away from the ISM absorption.
This is why astrospheric absorption is detectable despite being highly
blended with the ISM absorption (see Paper~2).  Thus, it is surprising
that astrospheric absorption has been detected for a star with such a
low $V_{ISM}$ value, and our astrospheric models have not been able
to explain the observed absorption.  More work must be done in the
future to offer a satisfactory explanation for this unusual case.

     In Figure~3a we plot our measured mass loss rates (per unit surface
area) versus X-ray surface fluxes.  For solar-like stars, X-ray emission
and winds both arise from the hot stellar coronae.  Thus, one might
expect to see a correlation of some sort between X-ray emission and mass
loss.  For the main sequence stars, mass loss appears to increase with
activity for $\log F_{X}<8\times 10^{5}$ ergs~cm$^{-2}$~s$^{-1}$.
The power law relation that we have fitted to the data in Figure~3a,
$\dot{M}\propto F_{x}^{1.34\pm 0.18}$,
is consistent with that reported in Paper~1.  However, the
new $\xi$~Boo data point suggests that the relation does not extend to
high activity levels.

     In Paper~1, we suggested that the apparent inconsistency of
Proxima~Cen (M5.5~V) and $\lambda$~And (G8~IV-III+M~V) with the
mass-loss/activity relation was due to these stars being less solar-like
than the GK main sequence stars, but the new low mass loss measurement
for $\xi$~Boo, which is a G8~V+K4~V binary, suggests that the relation
simply does not extend to high activity levels for {\em any} type of star.
Therefore, the power law in Figure~3a has been
truncated at $\log F_{X}=8\times 10^{5}$ ergs~cm$^{-2}$~s$^{-1}$.  All five
of the higher activity stars have mass loss rates much lower than the power
law would suggest.  The three evolved stars in Figure~3a ($\delta$~Eri,
$\lambda$~And, and DK~UMa) do not seem to have mass loss rates consistent
with those of the main sequence stars.  The very active coronae of
$\lambda$~And and DK~UMa produce surprisingly weak winds, though it should
be noted that both of these astrospheric detections are flagged as being
questionable in Paper~2.  Clearly more mass loss measurements would be
helpful to better define the mass-loss/activity relation of cool main
sequence stars, especially at high activity levels where more measurements
of truly solar-like G and K dwarf stars are necessary to see exactly what
is happening to solar-like winds at high coronal activity.  Additional
measurements are also required to determine whether G, K, and M dwarfs all
show the same mass-loss/activity relations.  Currently our sample is
simply not large enough to precisely address these questions.

     Why does the mass-loss/activity relation apparently change its
character at $\log F_{X}\approx 8\times 10^{5}$ ergs~cm$^{-2}$~s$^{-1}$?
One possible explanation concerns the appearance of
polar spots for very active stars.  Low activity stars presumably have
solar-like starspot patterns, where the spots are always
confined to low latitudes.  However, for very active stars not only are
spots detected at high latitudes, but a majority of these stars show
evidence for large polar spots \citep{kgs02}. The appearance of high
latitude and polar spots represents a fundamental change in the stellar
magnetic geometry \citep{cjs01}, and it is possible
that this dramatic change in the magnetic field structure could
affect the winds emanating from these stars.  We hypothesize that
stars with polar spots might have a magnetic field with a strong dipolar
component that could envelope the entire star and inhibit stellar
outflows, thereby explaining why active stars have weaker winds than
the mass-loss/activity relation of less active main sequence stars would
predict.  For $\xi$~Boo~A, high latitude spots of some sort have been
detected \citep{cgt88}, and \citet{pp05} have detected both global
dipole and large-scale toroidal field components.

     As we did in Paper~1, we combine the mass-loss/activity relation in
Figure~3a with a known relation between activity and age,
$F_{x}\propto t^{-1.74\pm 0.34}$ \citep{tra97}, to derive an
empirical mass loss evolution law for solar-like stars:
$\dot{M}\propto t^{-2.33\pm 0.55}$.  Figure~3b shows what this relation
suggests for the mass loss history of the Sun.  The truncation of the
power law relation in Figure~3a leads to the mass-loss/age relation in
Figure~3b being truncated as well at about $t=0.7$~Gyr.  Mass loss rates
for very active stars are significantly lower than would be predicted by
the mass-loss/activity relation defined by the less active stars, with the
$\xi$~Boo example being particularly relevant since the stars in this
binary are easily the most solar-like of those in this high activity regime.
Thus, the location of $\xi$~Boo is
shown in Figure~3b in order to infer what the solar wind might have been
like at times earlier than $t=0.7$~Gyr.

     The history of the solar wind is not only of interest to solar/stellar
astronomers, but it is also important for planetary studies \citep{ir05}.
Stellar winds can potentially erode planetary atmospheres, and the
strong winds that apparently exist for young stars make it even more likely
that winds have a significant impact on planets.  Work has begun on how
stellar winds might affect the atmospheres of detected extrasolar planets
\citep{jmg04}.  In our own solar system, the impact of the
solar wind on the atmospheres of Venus and Titan has been explored
\citep{ec97,hl00}, but the most intriguing case
study by far is Mars, since the history of the Martian atmosphere is
intimately connected with the history of water and perhaps life on the
surface of the planet.  Mars lost most its atmosphere early in its history,
possibly due to solar wind erosion \citep[e.g.,][]{hl03}.
This did not happen on Earth, presumably due to the protection from
the solar wind provided by the Earth's strong magnetosphere.  Unlike Earth,
Mars lost its global magnetic field at least 3.9 Gyr ago \citep{mha99},
and this is roughly the period when most of its atmosphere and
surface water are believed to have disappeared as well.  Interestingly
enough, the time when Mars is believed to have lost most of its atmosphere
corresponds roughly to the time when Figure~3b suggests that the solar wind
abruptly strengthens ($t\approx 0.7$ Gyr).  Perhaps this
strengthening of the solar wind, which we have speculated might be
connected to the loss of polar spots, played a central role in the
dissipation of the Martian atmosphere.

\acknowledgments

     Support for this work was provided by NASA grant NNG05GD69G and
grant AR-09957 from STScI.

\clearpage

\begin{table}[b]
\begin{center}
Table 1.  Mass Loss Measurements from Astrospheric Detections
\scriptsize
\begin{tabular}{lccccccc} \hline
Star & Spectral & $d$ & $V_{ISM}$ & $\theta$ & $\dot{M}$ & Log L$_{x}$ &
  Surf.\ Area \\
 & Type & (pc) & (km~s$^{-1}$) & (deg) & ($\dot{M}_{\odot}$) & &
  (A$_{\odot}$) \\
\hline
\multicolumn{8}{c}{PREVIOUS ANALYSES} \\
\hline 
Proxima Cen   & M5.5 V      & 1.30 & 25 & 79 &$<0.2$& 27.22 & 0.023\\
$\alpha$ Cen  & G2 V+K0 V   & 1.35 & 25 & 79 & 2    & 27.70 & 2.22 \\
$\epsilon$ Eri& K1 V        & 3.22 & 27 & 76 & 30   & 28.32 & 0.61 \\
61 Cyg A      & K5 V        & 3.48 & 86 & 46 & 0.5  & 27.45 & 0.46 \\
$\epsilon$ Ind& K5 V        & 3.63 & 68 & 64 & 0.5  & 27.39 & 0.56 \\
36 Oph        & K1 V+K1 V   & 5.99 & 40 &134 & 15   & 28.34 & 0.88 \\
$\lambda$ And & G8 IV-III+M V&25.8 & 53 & 89 &  5   & 30.82 & 54.8 \\
\hline 
\multicolumn{8}{c}{NEW ANALYSES} \\
\hline 
EV Lac        & M3.5 V      & 5.05 & 45 & 84 &  1   & 28.99 & 0.123\\
70 Oph        & K0 V+K5 V   & 5.09 & 37 &120 & 100  & 28.49 & 1.32 \\
$\xi$ Boo     & G8 V+K4 V   & 6.70 & 32 &131 &  5   & 28.90 & 1.00 \\
61 Vir        & G5 V        & 8.53 & 51 & 98 & 0.3  & 26.87 & 1.00 \\
$\delta$~Eri  & K0 IV       & 9.04 & 37 & 41 &  4   & 27.05 & 6.66 \\
HD 128987     & G6 V        & 23.6 &  8 & 79 &  ?   & 28.60 & 0.71 \\
DK UMa        & G4 III-IV   & 32.4 & 43 & 32 & 0.15 & 30.36 & 19.4 \\
\hline
\end{tabular}
\normalsize
\end{center}
\end{table}

\clearpage

\begin{figure}[b]
\plotfiddle{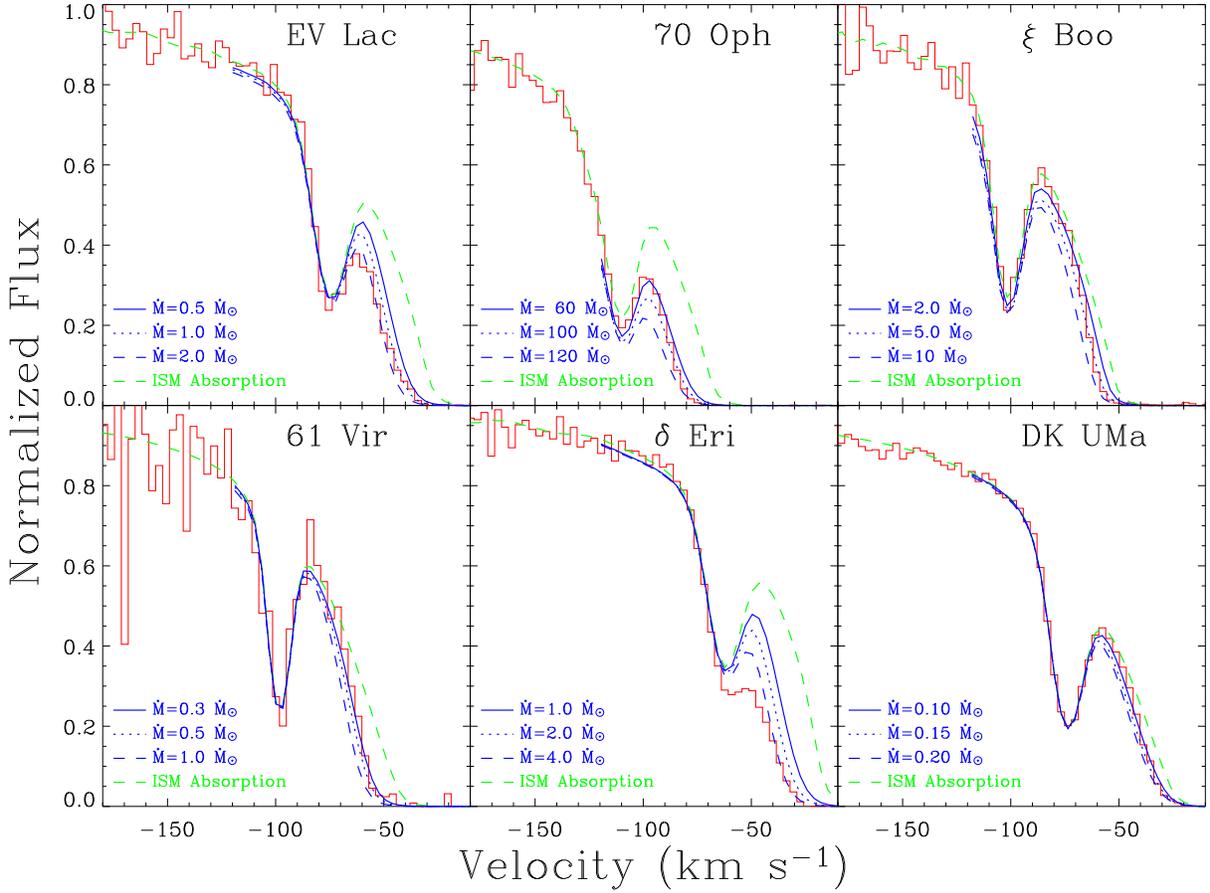}{3.3in}{90}{75}{75}{285}{-40}
\caption{The blue side of the H~I Ly$\alpha$ absorption lines for
  the new astrospheric absorption detections.  The green dashed lines show
  the ISM absorption, and the blue dashed lines show the additional
  astrospheric absorption predicted by models assuming various mass loss
  rates.}
\end{figure}

\clearpage

\begin{figure}
\plotfiddle{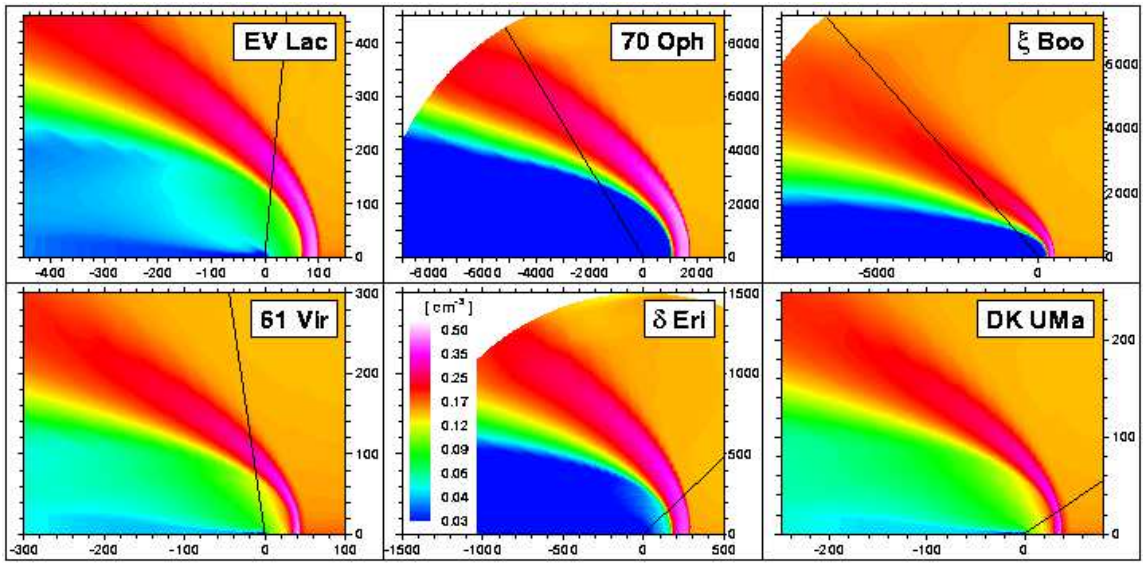}{3.3in}{-90}{60}{60}{-245}{300}
\caption{Maps of H~I density for the astrospheric models that
  yield the best fits to the data in Fig.~1.  The distance scale is in
  AU.  The laminar ISM wind seen by the star comes from the right.
  Solid lines indicate the observed Sun-star line of sight.}
\end{figure}

\clearpage

\begin{figure}[h]
\plotfiddle{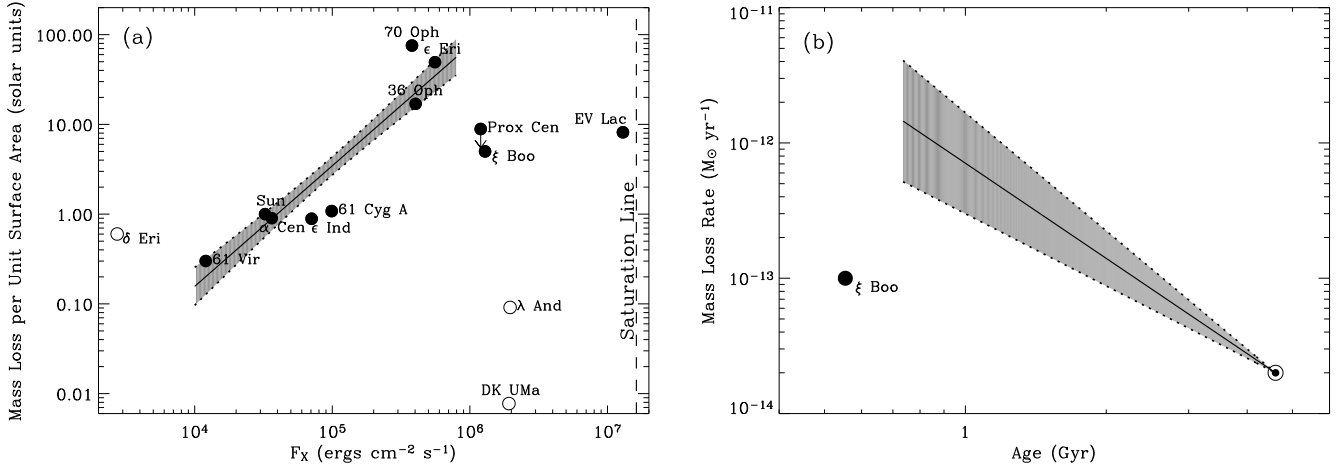}{3.5in}{90}{75}{75}{300}{-180}
\caption{(a) Mass loss rates per unit surface area plotted versus stellar
  X-ray surface fluxes.  Filled symbols are for main sequence stars,
  while open symbols are for evolved stars.  For main sequence stars
  with $\log F_{X}<8\times 10^{5}$ ergs~cm$^{-2}$~s$^{-1}$, mass loss
  increases with X-ray flux and we have fitted a power law to these data
  points.  Uncertainties in this relation are estimated as described in
  Wood et al.\ (2002).  (b) The mass loss history of the Sun inferred from
  the power law relation in (a).  The truncation of the relation in (a)
  means that the mass-loss/age relation is truncated as well.  The low mass
  loss measurement for $\xi$~Boo suggests that the wind suddenly weakens
  at $t\approx 0.7$ Gyr as one goes back in time.}
\end{figure}

\end{document}